# *Ab-initio* investigation of hot electron transfer in CO$_2$ plasmonic photocatalysis in presence of hydroxyl adsorbate

Zelio Fusco[1], Dirk Koenig[2], Sean C. Smith[2], Fiona Jean Beck[1]
[1]Renewable fuel group, School of Engineering, College of Engineering, Computing and Cybernetics, The Australian National University, Canberra, ACT 2601, Australia
[2]Integrated Materials Design Lab, The Australian National University, Canberra, ACT 2601, Australia

**Abstract**
Photoreduction of carbon dioxide (CO$_2$) on plasmonic structures is of great interest in photocatalysis to aid selectivity. While species commonly found in reaction environments and associated intermediates can steer the reaction down different pathways by altering the potential energy landscape of the system, they are often not addressed when designing efficient plasmonic catalysts. Here, we perform an atomistic study of the effect of the hydroxyl group (OH) on CO$_2$ activation and hot electron generation and transfer using first-principle calculations. We show that the presence of OH is essential in breaking the linear symmetry of CO$_2$, which leads to a charge redistribution and a decrease in the O=C=O angle to 134°, thereby activating CO$_2$. Analysis of the partial density of states (pDOS) demonstrates that the OH group mediates the orbital hybridization between Au and CO$_2$ resulting in more accessible states, thus facilitating charge transfer. By employing time-dependent density functional theory (TDDFT), we quantify the fraction of hot electrons directly generated into hybridized molecular states at resonance, demonstrating a broader energy distribution and a 11% increase in charge-transfer in the presence of OH groups. We further show that the spectral overlap between excitation energy and plasmon resonance plays a critical role in efficiently modulating electron transfer processes. These findings contribute to the mechanistic understanding of plasmon-mediated reactions and demonstrate the importance of co-adsorbed species in tailoring the electron transfer processes, opening new avenues for enhancing selectivity.

**Introduction**
Subwavelength metallic nanoparticles that support plasmonic resonances represent a versatile platform for nanoscale light manipulation. These structures hold great potential in nanophotonic applications that benefit from enhanced near-field light-matter interactions, such as plasmon-enhanced biosensing[1-3], surface-enhanced Raman scattering[4, 5] and photocatalysis[6-8]. Harnessing the hot electrons (HE)[9] generated by the non-radiative decay of illuminated metal nanoparticles before they thermalise is an attractive way to initiate and selectively drive chemical reactions[10-13], and has recently attracted significant interest in the heterogeneous catalysis community[14-18]. Hot electrons can be rapidly transferred to orbitals of nearby molecules, thereby driving catalytic chemical reactions. The energy and momentum distribution of the hot carriers depend on multiple factors, including material, particle morphology, crystal orientation, defects and environs[19, 20], influencing many aspects of heterogeneous catalysis, including reaction rate, selectivity, turnover frequency, yield, and reaction order[21]. To date, the significant role of HEs provided by photoexcited plasmons have been attributed to many reactions, including water splitting[22], bond cleavage[23] or formation[24], and carbon dioxide reduction[25].

Once generated, HEs can be transferred to nearby molecules directly or indirectly[26, 27]. In the former process, HEs are directly excited from the metal into unoccupied molecular orbitals of interacting acceptor adsorbates. The latter process comprises HEs generated within metal nanoparticles, subsequentially being scattered to empty states of surface-adsorbed molecules with suitable overlaps in energy and momentum distribution. The direct injection of HEs is a faster and more efficient process compared to the indirect transfer[28], and offers great potential for enabling high selectivities to adsorbed chemical species and their reaction paths[28-30]. High selectivities can be achieved by precisely tuning the plasmon resonance and the energy distribution of HEs to align with targeted unoccupied adsorbate states, thereby selectively promoting a specific reaction pathway[6, 17, 30]. The design and engineering of the optoelectronic properties of plasmonic architectures is thus important in particular to achieve selectivity in multiproduct reactions such as $CO_2$ reduction[31].

In the context of plasmon-enabled chemical transformations, reactions pathways and activation energies are often being investigated with ground-state density functional theory (DFT) methods[32-35]. Although these analyses provide important information on the systems and contribute to the understanding of the action mechanisms, they cannot accurately capture the physics of excited-states and the dynamics of the photogenerated hot carriers. Instead, time-dependent DFT (TDDFT) approaches have been increasingly employed to shed light on plasmon dynamics and hot carrier generation[36-38] by providing atomic-scale insights into excitation and transfer processes of electrons. In particular Rossi et al. have developed a real-time TDDFT methodology to study the dynamics of plasmon-molecules systems[39-43], demonstrating that the generated HEs have an energetic and spatial distribution that depends on the atomic structure, with lower-coordinated atoms exhibiting a higher proportion of HEs[41]. These can be directly transferred to adjacent semiconductors[40] or molecules[29, 42], even when the molecule is not chemisorbed to the plasmonic nanoparticle[42]: using Ag-CO as a model system, Fojt *et al.* demonstrated that HEs can be directly generated on the molecule at distances up to 5Å.

We extend the approach of Rossi et al. to investigate the energy required to activate carbon dioxide ($CO_2$) on a small Au cluster in the presence of co-adsorbed environmental species and probe the carrier dynamics. It is widely accepted that the reaction environment and intermediate species can steer the reaction pathway[44-46] by altering the potential energy landscape of the chemical transformation. However, a mechanistic study of the effect of intermediate adsorbates on the HE generation and transfer in $CO_2$ reduction is still missing. Here, we focus specifically on the role of adsorbed hydroxyl groups, as they take part in the $CO_2RR$ and are crucial for enhanced stability, activity and selectivity[47-49]. By using TDDFT calculations, we study the HE transfer process as a function of the distance, excitation frequency and an increasingly non-monochromatic frequency distribution, extending the fundamental understanding of HE transfer across nanoparticle-molecule interfaces and their impact on the overall $CO_2$ reduction reaction.

**Results and Discussion**
**System description**
We consider $CO_2$ as a reactant species because its reduction is of significant industrial and societal importance: artificial synthesis of hydrocarbons and alcohols from $CO_2$, using water to provide hydrogen and sunlight to drive the reaction, has the potential to provide renewable fuels and chemicals at scale[17, 50]. A considerable challenge in converting $CO_2$ into organic products lies in its activation, due to the notably high reaction barrier linked to the initial electron transfer (-1.9V vs standard hydrogen electrode). Adsorption of $CO_2$ on catalytic surfaces forms partially charged $CO_2^{\delta-}$ species which –due to charge redistribution– results in a change in the O=C=O bond angle and leads to a bent configuration[51, 52], as shown in Fig. 1a. As a consequence, the lowest unoccupied molecular orbital (LUMO) shifts to lower energies, reducing the energy barrier required for electron transfer and promoting a $CO_2$ reduction reaction[53].

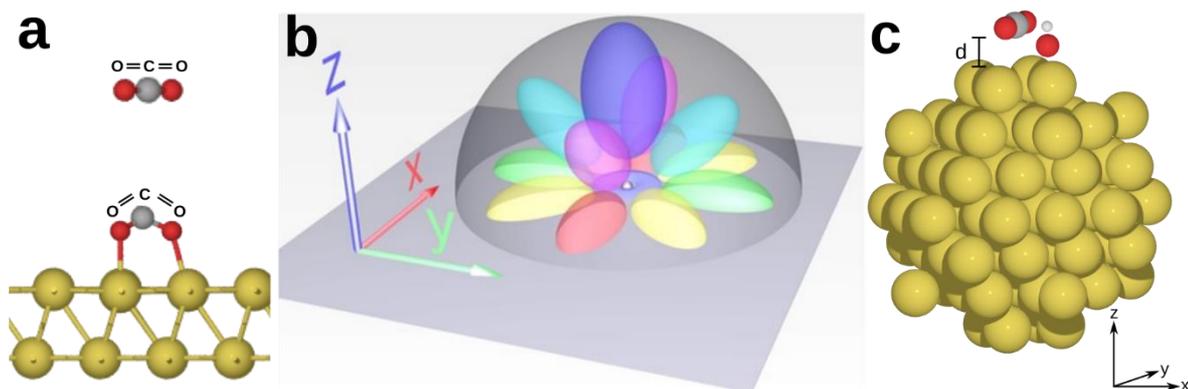

*Figure 1. a) Activation of CO₂ on a catalytic surface. The free molecule has a linear configuration with a O=C=O bond angle of 180°; upon interaction with the surface atoms the bond angle reduces, thus activating $CO_2$. b) Spatial scheme of the valence orbitals of a gold atom (bright grey sphere) on the catalytic surface: Colour-coding represents the spherical 6s (grey), $d_{xy}$ (yellow), $d_{x^2-y^2}$ (red and green), $d_{xz}$ (purple), $d_{yz}$ (cyan) and $d_{z^2}$ (blue). Colours refer to AO lobes aligned to x-/y-/z-axis (red/green/blue), and superimposed colours for bi-axial orientation; xy shown by yellow, xz shown by magenta, and yz shown by cyan. c) Schematic representation of the system analysed in this work, comprising of a $Au_{87}$ cluster with an OH group and $CO_2$.*

We use gold as a standard plasmonic material because of its unique electronic configuration which allows it to efficiently mediate catalytic reactions[31, 54]. The electronic configuration of Au is shaped considerably by its atomic orbital arrangement. Gold has a $5d^{10}6s^1$ valence orbital configuration which consists of five filled 5d atomic orbitals that enable hybridization with various reactant molecules, and a partially filled 6s orbital. The five 5d orbitals are strongly lobed and are overlayed by the diffused spherical 6s orbital which does not hybridize with the 5d orbitals. The large nuclear charge of gold leads to a more pronounced nuclear-electron attraction of the valence electron shell: due to relativistic effects, the s-orbitals are found to contract in response to the large nuclear charge, while the d-orbitals are in fact expanded and thus increases their involvement in the chemical and physical properties of gold[55, 56]. These specific atomic orbital lobes allow for various combinations of hybridisation and complex formation with molecular species and ligands, making gold a promising material for catalytic applications and HE transfer. It is worth noting that this versatility may pose challenges to spatial selectivity, as gold can couple with numerous molecular species. However, the resulting higher spatial probability of an electron transfer due to an increased overlap integral of the atomic and/or molecular orbitals involved promotes catalytic reactions. Selectivity for a specific reaction can be achieved by carefully tuning and aligning the energy distribution of HEs with the energy of the lowest unoccupied molecular orbitals (LUMO) of the reactant species[28], as discussed at the end of this section.

**Electronic ground state properties**
We start by investigating the adsorption and activation of a $CO_2$ molecule on a $Au_{87}$ cluster in presence and absence of an adsorbed hydroxyl group by performing a structural relaxation. The two model systems are $Au_{87}$ + $CO_2$ and $Au_{87}$ + OH + $CO_2$, with the $CO_2$ initially placed at a varying distance 2 Å < $d$ < 5 Å on a bridge-site, as this is the most catalytically active site for $CO_2$ reduction on gold surfaces[57], while the OH group is adsorbed on an on-top site.

Figure 2a shows the O=C=O bond angle when the systems reached equilibrium (forces < 0.05 eV/atom). In absence of the OH group, $CO_2$ maintains a linear symmetry with a final equilibrium O=C=O angle which fluctuates around 178°, independent of the distance from the Au cluster (red curve). Contrarily, the presence of OH group perturbs the charge symmetry of the molecule, leading to a structure transformation for $d$ < 3 Å. This results in a bent configuration with a final equilibrium O=C=O angle of 133°, in line with previous theoretical reports on activated $CO_2$ on catalysts surfaces[51, 52]. Similar results are also obtained on periodic gold slabs (see Supporting Information

S1). Simultaneously, as the CO$_2$ approaches the Au cluster ($d$ < 3 Å), a redistribution of the charge density occurs, as can be seen by the different LUMO shapes of the system (Fig. 2b). These results suggest the importance of investigating the effects of co-adsorbed hydroxyl groups – and possibly other environmental species – in the overall activation of CO$_2$, as well as the role of such species in HE transfer for efficiently driving catalytic processes.

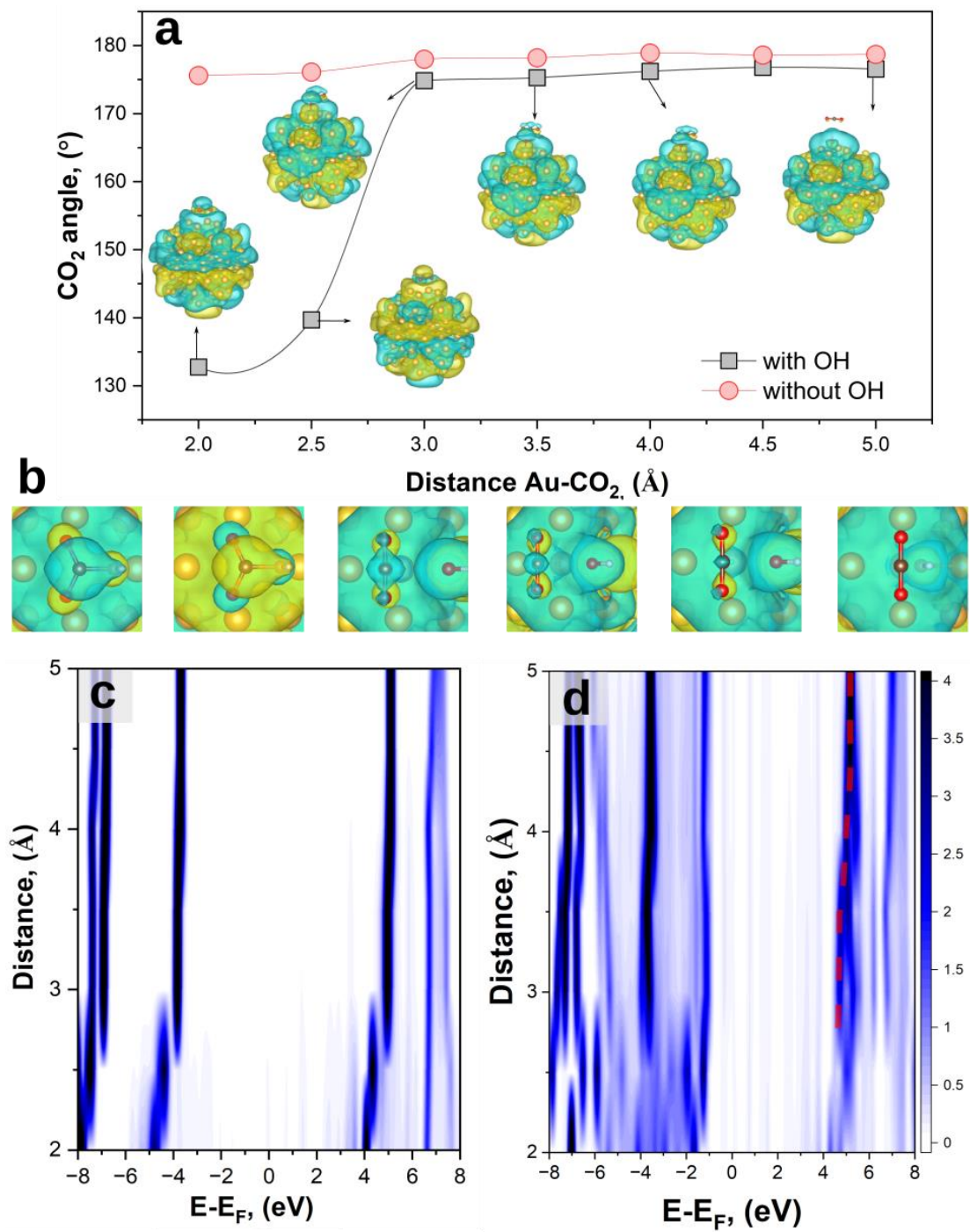

*Figure 2. a) CO$_2$ bond angle as a function of the distance during the adsorption process on Au and Au + OH clusters, red and black curves respectively. The LUMO wave functions of the systems are also shown at a the same isosurface value of 0.05 a.u. b) Magnification of the top-view of the LUMO at the same distances of (a). At a d = 2.5 Å, when transitioning from weak to strong hybridization (from 3 Å to 2.5 Å), we observe a phase flip in the molecular orbital (MO) lobes. This is due to the spin eigenfunction of the MO changing state, i.e. from spin-up to spin-down. When a phase flip happens, it means that the energy difference between both spin states of the MO is very small (≤ $k_BT$) so that both electrons can swap their spin eigenfunction (spinor) within the MO. The spin-flip occurrence when going from 3 Å to 2.5 Å and then back to the original spin configuration shows that one spin configuration of the LUMO has a slightly higher $E_{LUMO}$ compared to the $E_{LUMO}$ at d = 2 Å. c, d) Map of the projected density of states as a function of the distance and energy for Au + CO$_2$ and Au + CO$_2$ + OH.*

When $CO_2$ approaches the Au and Au + OH clusters, the strength of orbital hybridization increases[42]. This can be demonstrated by analysing the projected density of states (pDOS) of the adsorbed molecules as a function of their distance to the Au surface for the different systems (Fig. 2c, d). At distances d ≳ 4 Å for both, Au+$CO_2$ and Au+OH+$CO_2$ systems, the pDOS converges to the molecular density of states (DOS) of the isolated molecules. As *d* decreases, the pDOS eventually splits into several branches due to hybridisation with Au orbitals[42]. The addition of the OH group qualitatively shows a similar trend with a branched pDOS (Fig. 2b), but with an increased number of electronic states. In other words, the OH group mediates the orbital hybridization between Au and $CO_2$, allowing for more accessible states.

We anticipate that these states are of significant importance for the further HE transfer process[31]; in fact, a HE transfer to the adsorbed molecule requires the presence of acceptor states at suitable energies, i.e. where the pDOS is present[42], thus having more unoccupied levels which can lead to an increase of catalytic efficiency.
Additionally, this adsorption process is also accompanied by a shift towards a lower binding energy of the LUMO, decreasing from ~4.9 eV (*d* ≳ 4 Å) to ~4.6 eV (*d* ~ 3.5 Å) in line with previous literature[14], eventually splitting up into 4.7 eV (LUMO+1) and 4.5 eV (LUMO) for *d* < 3 Å due to exchange interactions rising with inverse inter-atomic distance (with the overlap integral of associated orbitals forming a bond). This is another indication that the OH group acts as a "bridge" between $CO_2$ and Au in terms of electronic coupling, which can then enable electron transfer from a considerably larger distance.

The total DOS of the different systems is shown in Figure S2. The introduction of $CO_2$ and OH induces only minor alterations in the total DOS due to the dominant contribution of states originating from the Au cluster. Notably, in the presence of the molecules, the Au cluster exhibits a slightly positive charge, which becomes more pronounced when the OH group is involved. The shift of the HOMO to higher energy is ~0.1 eV for Au+$CO_2$ (red curve) and ~0.16 eV for Au+$CO_2$+OH (green curve). This indicates that both the adsorbed OH and the $CO_2$ attract negative charges from the Au nanoparticle, as becomes apparent when comparing electronegativities, ionization energies, and electron affinities of C and O vs. Au [58]. Consequently, an electron transfer should occur from the Au NP to the molecular species. The $CO_2$ molecule induces a significantly increased shift of the HOMO to higher binding energy as compared to the shift when the OH group is included in the system. This dominant energy shift due to the $CO_2$ molecule suggests that an electron transfer from the Au cluster will eventually reside with the $CO_2$ molecule rather than with the OH group, the latter presenting merely a transient charge path for transferring an electron to the $CO_2$ molecule.

**Plasmon dynamics and hot electron (HE) generation**
We now study the time evolution of the Au, Au + $CO_2$ and Au + $CO_2$ systems under light illumination; when present, $CO_2$ is placed at a distance of 2 Å from the Au surface. First, the photo absorption spectra of the system are calculated by applying an initial Dirac pulse (δ-kick) in the linear-response regime along the z-direction (and perpendicular to the $CO_2$ molecule, when present). As the Fourier transform of the δ-kick contains all frequencies with equal proportion, this kick equally excites all the optically allowed transitions and allows us to extract the photoabsorption spectrum. Next, we analyse the energy evolution of the systems after excitation with ultra-short optical pulses at the plasmonic resonance energy, which allows us to calculate the energy distribution of hot carriers generated by the decay of the plasmon. This analysis allows us to estimate the plasmon dephasing time after which most of the energy has been transferred to hot carriers.

The photoabsorption spectra (Fig 3a) are characterized by a smoothly growing intensity starting at around 1 eV, with a plasmon peak at ~2.5 eV, followed by a wide modulation up to 4 eV. As expected, the plasmon peak is relatively weak, in line with previous theoretical works on small gold clusters[59-61] which showed that a clear plasmon emerges only when the cluster size exceeds 2 nm[62]. For the combined systems, we observe a slight red-shift (green and red curves in Fig. 3a) of the plasmon peak from 2.66 eV for the single Au cluster, to 2.56 eV and 2.52 eV for Au + $CO_2$ Au + $CO_2$ + OH, respectively. This indicates the presence of additional decay channels for the plasmon due to formation of hybridized cluster-molecule states.

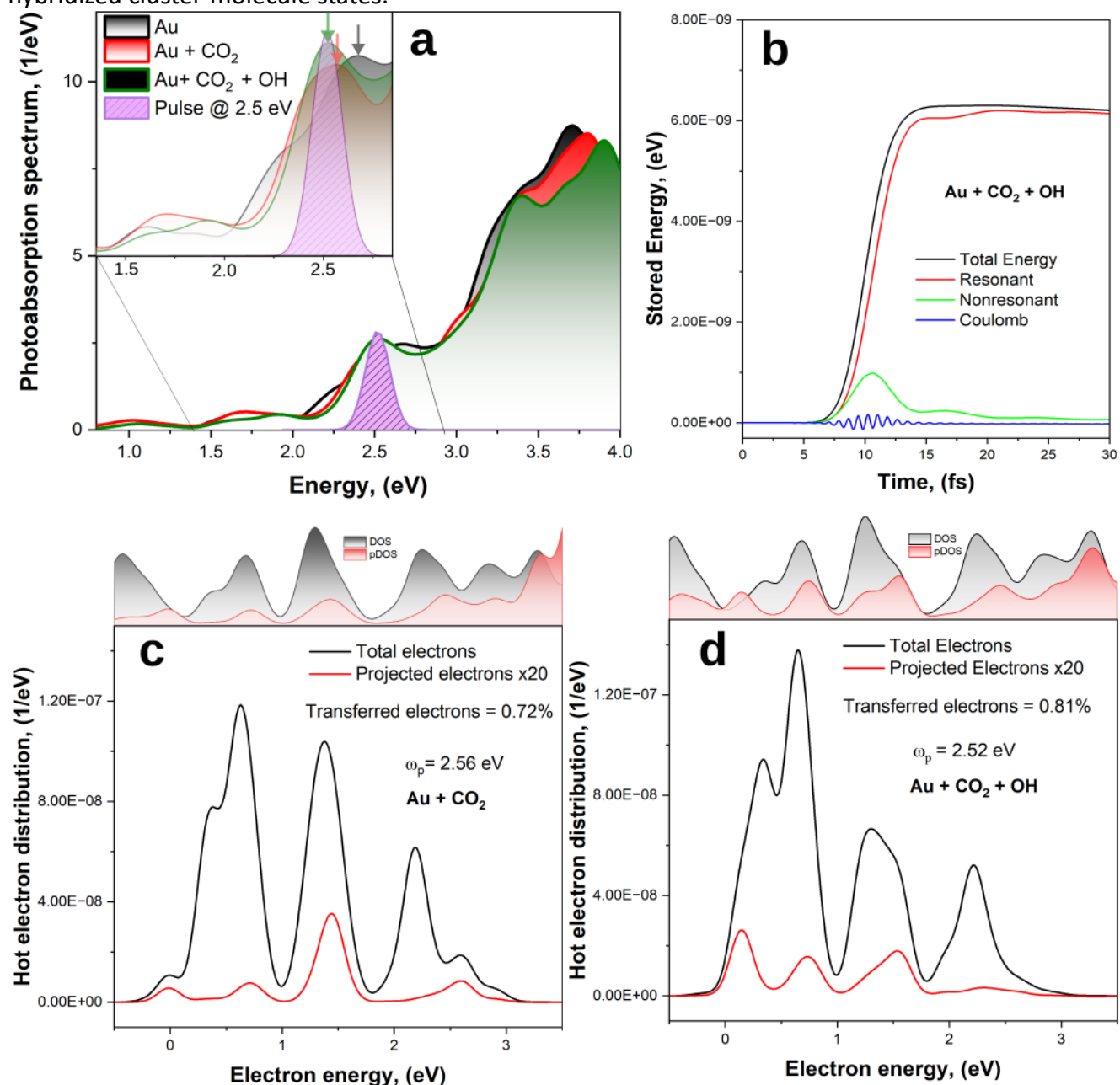

Figure 3. a) Photoabsorption spectra of Au, Au + $CO_2$ and Au + $CO_2$ + OH. The purple curve represents a typical driving laser pulse in the frequency domain, centered at the LSPR of the Au + CO2 + OH. The colour coded arrows represent the LSPR of the different systems. b) Time evolution of the stored energy for Au + $CO_2$ + OH, highlighting the different energy contributions. c-d) Generated and transferred HEs energy distribution for the Au + $CO_2$ and Au + $CO_2$ + OH systems, respectively, averaged between 15 and 30 fs.

Next, we excite the systems with a Gaussian pulse – $E(t) = E_0 \cos(\omega_p(t - t_0))\exp(-\sigma^2(t - t_0)^2/2)$ – with $\omega_p$ set to be in resonance with the systems ($\omega_p$ = 2.66 eV for Au, 2.56 eV for Au + $CO_2$ and 2.52 eV for Au + $CO_2$ + OH), centred at 10 fs and with a temporal FWHM of 5 fs (Fig3 a, purple peak).

The energy of the light pulse is absorbed by the electrons in the system, promoting them into excited states. The electronic energy is not equally distributed, and can be divided into Coulomb energy, and

electron–hole transition energy contributions[41, 63] as shown for the Au + OH + $CO_2$ system in Fig. 3b. The latter can be further separated into the energy of transitions resonant with the excitation pulse (i.e. with frequency ω = $ω_p$ ± 2σ), constituting hot carriers, and non-resonant transitions, attributed to transitions from d-states in the metal nanoparticle which screen the plasmon. The evolution of the different energy contributions is similar for all the investigated systems (Fig S3), as the plasmon dephasing process is not affected on a significant scale by a single molecular species. Initially, the plasmon is excited via Coulomb interactions[64] and non-resonant contributions carry most of the energy during the plasmon excitation. After ~15 fs, as the plasmon decays, the total energy is redistributed into electron–hole excitations that are resonant with the pulse.

Having established that after ~15 fs the energy of the plasmon is mostly stored in hot carriers, we proceed to analyse the energy distribution of the photogenerated HEs, as well as the fraction of such HEs directly excited to states localised on the $CO_2$ molecule for the Au + $CO_2$ and Au + $CO_2$ + OH systems. The HE energy distributions are shown in Fig. 3c-d, where we have averaged the energies of the HEs between 15 fs and 30 fs as the resonant transitions have already reached their steady state. Therefore, HEs possess one temperature in accord with Fermi-Dirac statistics relative to the energy of their steady state as given by the convolution of the initial HE population with the total DOS of the system in the relevant energy range. For both the analysed systems, the generated HEs (black lines) can have energies up to the laser pulse frequency ($E_F$ + $\hbar ω_p$) due to the energy conservation. In addition, their energy distributions are not uniform within this range but show occupation probabilities that match the total DOS of the systems; electrons can only be excited to existing unoccupied states. The total DOS for Au + $CO_2$ and Au + $CO_2$ + OH are similar as they are dominated by the Au contributions. It follows that all systems have similar total HE energy distributions, presenting major peaks at ~0.7 eV, 1.3 eV and 2.2 eV.

Conversely, the transferred electron energy distributions (red lines), which represent the fraction of HEs directly excited to states localised on the $CO_2$, mirror the pDOS and exhibit differences between the systems. Specifically, in absence of the OH group, the fraction of HEs generated on the $CO_2$ is 0.72%, with most of them having an energy of 1.4 eV. The addition of the OH group increases the density of electronic states (see also Fig. 2b,) thus opening more channels for electron transfer and effectively improving the charge transfer to 0.81%, which corresponds to an 11% increase with respect to the case without OH.

In the next section, we investigate the effect of the excitation parameters –pulse frequency ($ω_p$) and width (FWHM)– on the direct excitation of HEs to $CO_2$ orbitals, by varying $ω_p$ from 1.6 eV to 2.8 eV, and the FWHM from 4 fs (1.03 eV) to 15 fs (0.28 eV). The energy distributions of the HEs excited to states localised on the $CO_2$ molecule are reported in Fig. 4a-d as a function of pulse energy; the distributions for the total amount of HEs in both systems are given in Fig. S4. The amount of generated HEs increases with the increase of pulse width. For FWHM = 4 fs, we obtain a maximum HE density when $ω_p$ ~ 2.8 eV, while for FWHM = 15 fs the maximum occurs when 2.3 < $ω_p$ < 2.7 eV for all systems. In this energy range, the pulse excitation approaches the plasmon resonance of the systems (Fig. 3a), thus emphasizing the critical role of spectral overlap in efficiently modulating electron transfer processes[65, 66]. Furthermore, below ~2 eV, photoabsorption decreases, thus resulting in a negligible generation of HEs within the molecule. This finding suggests that engineering the photoabsorption, i.e. the plasmon resonance energy, should be considered to efficiently leverage photogenerated HEs for tailored reactions.

In the absence of the OH group (Fig. 4a,c), the energy distribution of HEs exhibits a prominent, dominant single peak centred around ~1.4 eV for pulse energies of 2.4-2.6 eV. Notably, approximately

55% (75%) of the HEs assume this specific energy level when subjected to pulse durations with FWHM of 4 fs (15 fs) and under resonance conditions. Conversely, the presence of the OH group results in a broader electron energy distribution. This is attributed to the presence of more accessible hybridized states within the energy range of 0 eV – 2 eV, which can be populated by HEs. The broadened energy distribution as facilitated by the OH group provides more channels for HEs transfer and promotes the system to a different excited level, potentially leading to a different product distribution[6]. This demonstrates that the presence of the OH group in the $CO_2$ photoreduction process alters the energy landscape and transfer behaviour of HEs, thereby possibly affecting the reaction pathway, and highlight the importance of taking into account the influence of environmental and intermediate species for selective HE catalysis.

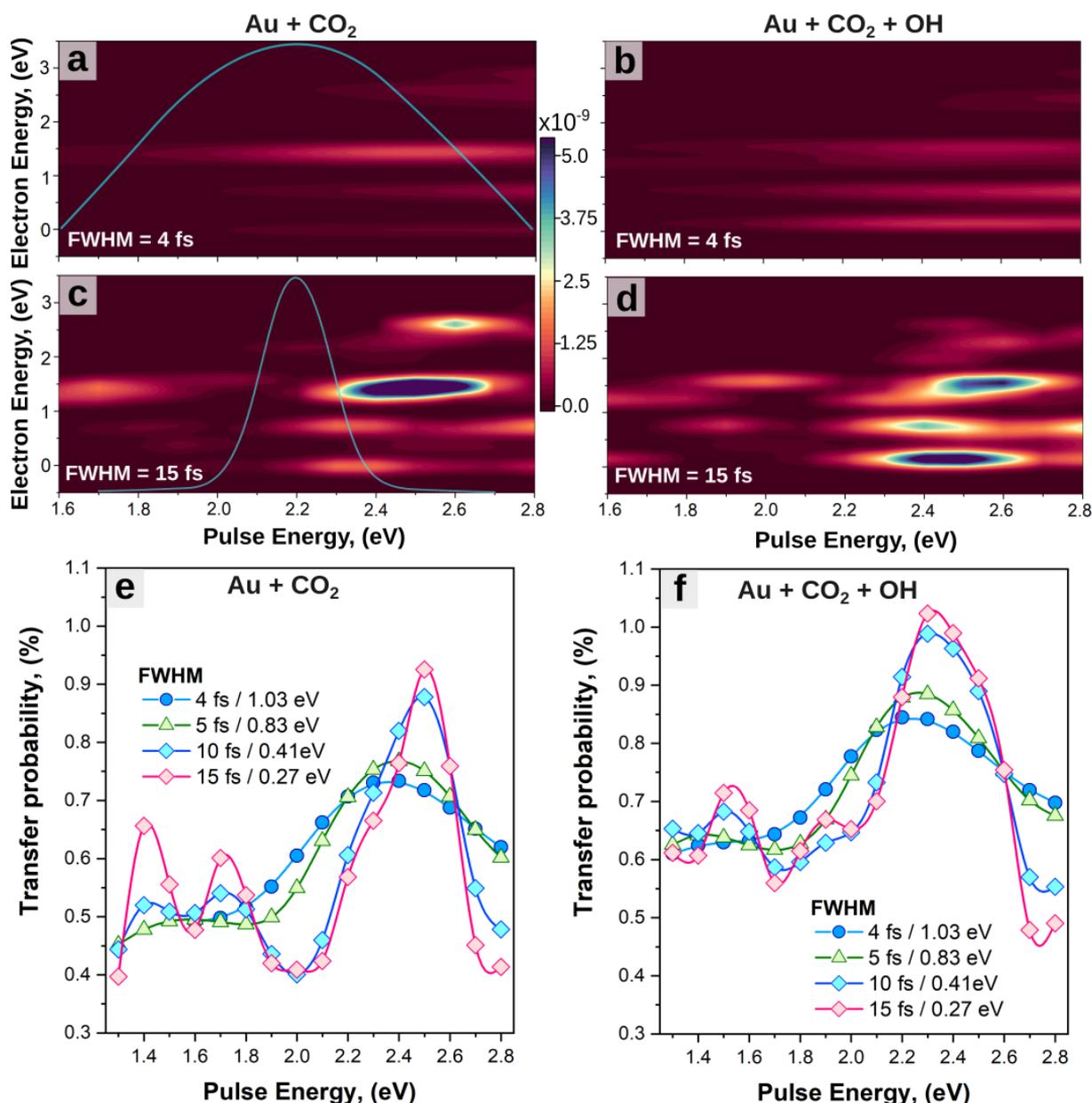

*Figure 4. a, b) Amount of HEs generated on $CO_2$ and their energy distribution as a function of the pulse frequency (with temporal FWHM of 4 fs (1.03 eV)) for Au + $CO_2$ and Au + $CO_2$ + OH, respectively. c-d) Same as a-b, but the excitation has a FWHM of 15 fs (0.27 eV). The light blue lines in a) and c) represent a typical laser pulse in frequency domain. e-f) Fraction of carriers that are being transferred to the molecule as a function of the excitation frequency for multiple FWHM.*

Furthermore, the excitation width also affects the HE distributions. Narrow band pulses (FWHM = 15 fs; ΔE = 0.27 eV) lead to an overall increase in the generation of HEs on the molecule; conversely, with

'broadband' excitation (FWHM = 4 fs; ΔE = 1.03 eV) the pulse energy is redistributed to a wider range of possible transitions in the photoabsorption spectrum, leading to a broader but less intense (less resonant) electron transfer. The transfer probability is given in Fig. 4 3-f, calculated as the ratio of the number of HEs generated on the molecule to the total HEs in the systems, and exhibits consistently higher values when the OH group is present, indicating a more efficient transfer. Specifically, the addition of the OH group leads to an increase of ~10 - 15% in the transfer probability, depending on pulse frequencies and FWHMs. Interestingly, when transitioning from narrowband to broadband excitation (FWHM = 0.24 eV to 1.03 eV), despite a reduced transfer probability, the electron transfer to acceptor hybridized states remains high, with a maximum probability of ca. 0.84% for Au + $CO_2$ + OH at $\omega_p$ = 2.3 eV. This result suggests that HE transfer can still occur under broadband solar illumination with minor reductions of the yield, which is critical for practical photocatalysts operation.

We note that, while the quantitative results are specific to the systems considered, the fundamental physics elucidated here remains broadly applicable for other systems. Importantly, our findings highlight the critical role of the OH adsorbate in determining (i) the adsorption and activation of $CO_2$, (ii) the hybridization between Au and $CO_2$, and (iii) the amount of HEs directly transferred into hybridized states. Consequently, the presence of OH species can promote the $CO_2$ reduction reaction to a different potential energy surface, possibly leading to different product distributions and reaction speeds. It is therefore very important to consider adsorbed species during the design of plasmonic catalysts to achieve high efficiency and selectivity for specific reactions, particularly when dealing with plasmonic structures functionalized with ligands.

**Conclusions**
We investigate the mechanisms of plasmon-driven $CO_2$ reduction in the presence of a co-adsorbed hydroxyl group via real-time time-dependent density functional theory. Our findings show that the OH group induces a redistribution of charges in $CO_2$, leading to favourable absorption and activation on small Au clusters. Critically, adding an OH group significantly increased hybridization between molecular states of $CO_2$ and electronic states of Au as mediated by the OH group, enabling and improving charge transfer. This coupling is distance-dependent, with a maximum when the $CO_2$ is close to the Au cluster, and can be predicted by analysing the branching of the partial density of states of the molecules.
By applying TDDFT, we are able to study the carrier dynamics of these systems at an atomic level, whereby the quantification of the hot electron (HE) energy distribution is critical. While the total HE distribution is largely unaffected by the presence of the OH group, the additional hybridized states facilitate the transfer of HEs to the $CO_2$ molecule, thereby increasing the resonant direct HE transfer by 11%. Furthermore, our findings confirm the importance of spectral overlap between the excitation energy and the plasmon resonance, and show that direct HE transfer is only slightly decreased (~1.05% to 0.84%) when the excitation is changed from narrowband to broadband. This finding shows that a significant chemical reaction rate still occurs under broadband excitation such as solar light.
In summary, this investigation elucidates the role of adsorbed hydroxyl groups in the adsorption and activation of $CO_2$ on plasmonic surfaces. It highlights the importance of considering adsorbed molecules to enhance direct HE transfer, ultimately enabling the design of highly efficient and selective plasmonic photocatalysts with polychromatic excitation.